\documentclass[12pt]{article}
\title{Single eta  production in heavy quarkonia transitions
}
\author{ Yu.A.Simonov, A.I.Veselov\\
State Research
Center\\Institute of Theoretical and Experimental Physics, \\
Moscow, 117218 Russia}

\newcommand{\beq}{\begin{eqnarray}}
 \newcommand{\eeq}{\end{eqnarray}}
\newcommand{\be}{\begin{equation}}
 \newcommand{\ee}{\end{equation}}

\def\fun#1#2{\lower3.6pt\vbox{\baselineskip0pt\lineskip.9pt
\ialign{$\mathsurround=0pt#1\hfil ##\hfil$\crcr#2\crcr\sim\crcr}}}

\newcommand{{\SD}}{\rm SD}

\newcommand{{\Mc}}{\mathcal{M}}

\newcommand{\veP}{\mbox{\boldmath${\rm P}$}}
\newcommand{\vep}{\mbox{\boldmath${\rm p}$}}
\newcommand{\veq}{\mbox{\boldmath${\rm q}$}}

\newcommand{\vek}{\mbox{\boldmath${\rm k}$}}

\newcommand{\lan}{\langle}
\newcommand{\ran}{\rangle}

\begin{document}
\maketitle
\begin{abstract} The $\eta$  production in the $(n,n')$
bottomonium transitions $\Upsilon (n) \to \Upsilon (n') \eta, $ is
studied in the method used before  for dipion heavy  quarkonia
transitions. The widths  $\Gamma_\eta(n,n')$ are calculated
without fitting parameters for $n=2,3,4,5, n'=1$.Resulting
$\Gamma_\eta(4,1)$ is found to be large in agreement with recent
data.

\end{abstract}

\section{Introduction}
The $\eta$ and $\pi^0$ production in heavy quarkonia transitions
is attracting attention of experimentalists for a long time
\cite{1}. The first result refers to  the $\psi(2S) \to J/\psi
(1S) \eta$ process (to be denoted as $\psi (2,1)\eta$ in what
follows, similarly for $\Upsilon $) with
$\frac{\Gamma_\eta}{\Gamma_{tot}}= (3.09\pm 0.08) \%$ \cite{1}, $
\Gamma_{tot}=337\pm 13$ keV.

For the $\Upsilon (2,1)\eta$ and $\Upsilon (3,1)\eta$ transitions
only upper limits $B<2\cdot 10^{-3}$ and  $B<2.2\cdot 10^{-3}$
were  obtained in \cite{2}  and \cite{3} correspondingly and
preliminary results appeared recently in \cite{4},  $B(\Upsilon
(2,1)\eta) =(2.5\pm 0.7 \pm 0.5)10^{-4}$ and  $B(\Upsilon (2,1)
\pi^0) < 2.1\cdot 10^{-4}(90\%$ c.l.). On theoretical side in
\cite{5} small ratios of  widths \be
\frac{\Gamma(\Upsilon(2,1)\eta)}{\Gamma(\psi(2,1)\eta)} \cong
2.5\cdot 10^{-3} ~~ {\rm
and}~~\frac{\Gamma(\Upsilon(3,1)\eta)}{\Gamma(\psi(2,1)\eta)} =
1.3 \cdot 10^{-3}\label{1}\ee have been predicted, with the model
property   that the bottomonium yields of $\eta$ would be smaller
than those of charmonium;specifically  in the method of  \cite{6},
the width ratio is proportional  to
$O\left(\left(\frac{m_c}{m_b}\right)^2\right)\approx 0.1$, for a
discussion see also \cite{6,7}.

However recently \cite{8} new BaBar data have been published on
$\Upsilon (4,1)\eta$ with  the branching ratio \be B(\Upsilon
(4,1)\eta) =(1.96\pm 0.06 \pm 0.09)10^{-4}\label{2}\ee and \be
\frac{\Gamma(\Upsilon(4,1)\eta)}{\Gamma(\Upsilon
(4,1)\pi^+\pi^-)}= 2.41\pm 0.40\pm0.12.\label{3}\ee

This latter result  is very large, indeed the corresponding ratio
for $\psi (2,1) \eta$ transition is $\approx 0.2$ and theoretical
estimates  (\ref{1}) from \cite{5} for a similar ratio of
$\Upsilon (3,1) \eta/\pi\pi$ yield 0.015. All this suggests that
another mechanism can be at work in single $\eta$ production and
below we exploit the approach based on the  Field Correlator
Method (FCM) \cite{9} recently applied to  $\Upsilon(n,n')\pi\pi$
transitions with $n\leq 3$ in \cite{9,10}, $n\leq 4$ in \cite{11}
and $n=5$ in \cite{12,13}.

The method essentially expoits  the mechanism of Internal Loop
Radiation (ILR) with light quark loop  inside heavy quarkonium and
has two fundamental parameters -- mass vertices in chiral light
quark pair $q\bar q$ creation $M_{br} \approx f_\pi$ and pair
creation vertex without pseudoscalars $M_\omega \approx 2\omega$,
where $\omega (\omega_s)$ is the average energy of the light
(strange) quark in the $B(B_s)$ meson. Those are calculated with
relativistic Hamiltonian \cite{14} and considered as  fixed for
all types of transitions $\omega =0.587$ GeV, $\omega_s=0.639$
GeV.

Any process of heavy quarkonium transition with emission of any
number of Nambu-Goldstone (NG) mesons is considered  in ILR as
proceeding via intermediate states of $B\bar B, B\bar B^*+c.c.,
B_s \bar B_s$ etc. (or equivalently $D \bar D$ etc.) with NG
mesons emitted at vertices.

For one $\eta$ or $\pi^0$ emission one has diagrams shown in
Fig.1, where dashed line is for the NG meson. As shown in
\cite{9,10,11}, based on the chiral Lagrangian derived in
\cite{15}, the meson emission vertex has the structure \be
\mathcal{L}_{CDL} = - i \int d^4 x \bar \psi (x) M_{br} \hat U (x)
\psi (x)\label{4}\ee \be \hat U =\exp \left( i\gamma_5
\frac{\varphi_a\lambda_a}{f_\pi}\right), \varphi_a \lambda_a
=\sqrt{2} \left( \begin{array}{ccc} \frac{\eta}{\sqrt{6}}
+\frac{\pi^o}{\sqrt{2}},& \pi^+,& K^+\\
\pi^-,&\frac{\eta}{\sqrt{6}} -\frac{\pi^o}{\sqrt{2}}, &K^o\\
K^-, &\bar K^0,& -\frac{2\eta}{\sqrt{6}}
\end{array}\right),
\label{5}\ee
 \hspace{-2cm} \small{
\unitlength 1mm 
\linethickness{0.4pt}
\ifx\plotpoint\undefined\newsavebox{\plotpoint}\fi 
\begin{picture}(137,43.25)(0,0)
\put(40.75,31.75){\oval(29,7)[]} \put(110,32.38){\oval(30,6.75)[]}
\put(26,31.5){\circle*{2.12}} \put(54.75,32){\circle*{2.06}}
\put(95.25,33){\circle*{.5}} \put(95,32.5){\circle*{2}}
\put(124.5,32.75){\circle*{1.58}}
\put(40.5,31.63){\oval(33.5,19.25)[]}
\put(109.63,32.38){\oval(33.25,18.75)[]}
\put(25.68,31.68){\line(1,0){.983}}
\put(27.65,31.75){\line(1,0){.983}}
\put(29.61,31.81){\line(1,0){.983}}
\put(31.58,31.88){\line(1,0){.983}}
\put(33.55,31.95){\line(1,0){.983}}
\put(35.51,32.01){\line(1,0){.983}}
\put(37.48,32.08){\line(1,0){.983}}
\put(39.45,32.15){\line(1,0){.983}}
\put(14.25,35){\line(1,0){9.5}} \put(15,27){\line(1,0){9}}
\put(56.75,35.25){\line(1,0){9}}
\put(56.5,27.25){\line(1,0){9.25}}
\multiput(83.25,34.75)(1.21875,.03125){8}{\line(1,0){1.21875}}
\put(83.5,27.25){\line(1,0){9.75}}
\put(126.5,36){\line(1,0){10.5}}
\put(125.25,27.75){\line(1,0){11.25}}
\multiput(124.18,32.93)(.039216,.033497){17}{\line(1,0){.039216}}
\multiput(125.51,34.07)(.039216,.033497){17}{\line(1,0){.039216}}
\multiput(126.85,35.21)(.039216,.033497){17}{\line(1,0){.039216}}
\multiput(128.18,36.35)(.039216,.033497){17}{\line(1,0){.039216}}
\multiput(129.51,37.49)(.039216,.033497){17}{\line(1,0){.039216}}
\multiput(130.85,38.62)(.039216,.033497){17}{\line(1,0){.039216}}
\multiput(132.18,39.76)(.039216,.033497){17}{\line(1,0){.039216}}
\multiput(133.51,40.9)(.039216,.033497){17}{\line(1,0){.039216}}
\multiput(134.85,42.04)(.039216,.033497){17}{\line(1,0){.039216}}
\put(37.75,15.25){\makebox(0,0)[cc]{( a )}}
\put(104,17.25){\makebox(0,0)[cc]{( b )}}
\end{picture}}
\vspace{-0.5cm}

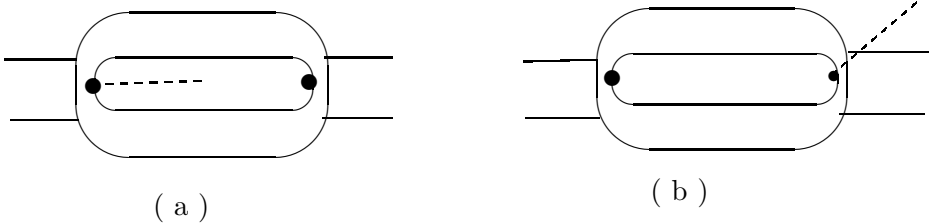
 Fig.1 Single eta
production (dashed line) from
 $\Upsilon(n)BB^*$ vertex (a), and
$BB^*\Upsilon(n')$ vertex (b).

\vspace{1cm}

The lines (1,2,3) in the $\hat U$ matrix  (\ref{2}) refer to
$u,d,s$ quarks and hence to the channels $B^+B^-, B^0\bar B^0,
B^0_s \bar B^0_s$ (and to the corresponding channels with $B^*$
instead of $B$). Therefore the emission of a single $\eta$  in
heavy quarkonia transitions requires the flavour $SU(3)$ violation
and resides  in our approach in the difference of channel
contribution $B\bar B^*$ and $B_s\bar B_s^*$, while the $\pi^0$
emission is due the difference of $B^0\bar B^{0*}$ and $B^+
B^{-*}$  channels (with $B\to D$ for charmonia).

The paper is devoted to the  explicit calculation of single $\eta$
 emission widths in bottomonium $\Upsilon (n,1)
\eta$ transitions with $n=2,3,4,5.$ Since theory has no fitting
parameters (the only ones, $M_\omega$ and $M_{br}$ are fixed by
dipion transitions) our predictions depend only on the overlap
matrix elements, containing wave functions of $\Upsilon (nS)$, $B,
B_s, B^*, B^*_s$. The latter have been computed previously in
relativistic Hamiltonian technic in \cite{14} and used extensively
in dipion transitions in \cite{11,12,13}.

The paper is organized as follows. In section 2 general
expressions for process amplitudes are given;  in section 3
results of calculations are presented and discussed and a short
summary and prospectives are given.

\section {General formalism}

The process of single NG boson emission in bottomonium transition
is described  by two diagrams depicted in Fig.1,  (a) and (b)
which can be written according to the  general formalism of FCM
\cite{9,11,12} as (we consider $\eta$ emission)

\be \Mc=\Mc_\eta^{(1)}+\Mc_\eta^{(2)}; \Mc_\eta^{(i)}=
\Mc^{(i)}_{B_sB^*_s} - \Mc^{(i)}_{BB^*} , i=1,2\label{6}\ee \be
\Mc_\eta^{(1)}=\gamma\int\frac{J^{(1)}_n (\vep, \vek) J_{n'}
(\vep)}{E-E(\vep)} \frac{d^3\vep}{(2\pi)^3},\label{7}\ee where
 $\Mc_\eta^{(2)}$ has the same form, but without  NG boson energy in the
 denominator of (\ref{7}). Here $
 \gamma=\frac{M_\omega M_{br}2}{\sqrt{2\omega_\eta} N_c f_\eta
 \sqrt{3}}$ and the overlap integral of $\Upsilon (nS)$ and $BB^*$
 wave functions is (for details see Appendix)

\be J_n^{(1)} (\vep, \vek) = \bar z^\eta_{123} ~^{(0)}I_{n,BB^*}
(p) e^{-\frac{p^2}{\Delta_n} -\frac{k^2}{4\beta^2_2}}\label{8}\ee

\be J_{n'} (\vep) = \bar z_{2} ~^{(1)}I_{n',BB^*} (p)
e^{-\frac{p^2}{\Delta_n'}}\label{9}\ee $\bar z_\eta(BB^*)$ and
$\bar z_2(BB^*)$ are Dirac traces of decay matrix elements
$\Upsilon(nS) \to BB^*\eta$ and $BB^*\to \Upsilon (n')$,
respectively  they are defined in \cite{9,11} and below in
Appendix. The special point  in our case is that $\eta$ meson is
emitted in $P$ wave, hence one must extract the corresponding term
in the Dirac trace, for details see Appendix.

\be \bar z^\eta (BB^*) \cdot \bar z_2 (BB^*) =\left(
\frac{M_b}{2\Omega}\right)^2\frac{4p^2u_ne_{i'il}}{3{\omega^3}}
k_l.\label{10}\ee

Here
$u_n=\frac{\beta^2_2\Omega}{\Delta_n(\omega+\Omega)}\approx\frac{\beta^2_2}{\Delta_n},$
and $\Omega(\Omega_s)$ is the average energy of the $b$ quark in
$B(B_s)$ meson; from Table IV in \cite{9}  one finds that
$\Omega=4.827$ GeV, $\Omega_s=4.830$ GeV. In what follows we shall
neglect the difference between $\Omega, \Omega_s$ and  the mass of
$b$ quark $M_b=4.8$ GeV. Note that these large masses cancel in
all matrix elements and final expressions will depend only on
energies $\omega$ and $\omega_s$ and differences of threshold
positions: $\Delta M{^*}=M(B)+ M(B^*) -M(\Upsilon (nS))$ and
$\Delta M_s$ -- the same for $B_s, B_s^*$ masses. Note, that the
contribution of the $B^*\bar B^*, B_s^*\bar B^*_s$ channels vanish
hence we shall consider only $B\bar B^*$ and $B_s\bar B^*_s$
channels.

Indices $i'i$ in $e_{i'il}$ in (\ref{10}) refer to the
$\Upsilon(n'S)$ and $\Upsilon (nS)$ polarizations respectively.
Finally,  coefficients $\beta_2,\beta_1$ and
$\Delta_n=2\beta^2_1+\beta^2_2$, refer to the expansion of
realistic wave functions of $\Upsilon(nS), \Upsilon(n'S)$ and
$B,B^*, B_s, B_s^*$  computed in \cite{14} in series of oscillator
functions and $\beta_1, \beta'_1$ and $\beta_2$ denote the
$\chi^2$ fitted oscillator parameters for those functions
respectively, see \cite{11} for details.

Finally we define  all quantities in the denominator of (\ref{7});
in $\Mc^{(1)}_{BB^*}$  the denominator is \be
E-E(\vep)=M(\Upsilon(nS))-(\omega_\eta+M_B+M^*_B+
\frac{\vep^2}{2M_B} +\frac{(\vep-\vek)^2}{2M^*_B})\equiv-\Delta
M^*- \omega_\eta-E(\vep, \vek).\label{11}\ee
\newcommand{{\Lc}}{\mathcal{L}}
For $\Mc_\eta^{(2)}$ one omits $\omega_\eta$ and $\vek$ in
(\ref{11}). Finally one can represent the matrix element
$\Mc^{(i)}_\eta$ as follows: \be \Mc_\eta^{(1)} =\gamma e_{ii'l}
k_l \frac{\beta^2_2}{3\Delta_n} \left(
\frac{1}{\omega^3_s}\Lc^{(1)}_s -\frac{1}{\omega^3} \Lc^{(1)}
\right) e^{-\frac{k^2}{4\beta^2_2}}\label{12}\ee

\be \Mc_\eta^{(2)} =\gamma e_{ii'l} k_l
\frac{\beta^2_2}{3\Delta_{n'}} \left(
\frac{1}{\omega^3_s}\Lc^{(2)}_s -\frac{1}{\omega^3} \Lc^{(2)}
\right) e^{-\frac{k^2}{4\beta^2_2}}\label{13}\ee with \be
\Lc^{(1)} =\int \frac{d^3p~p^2}{(2\pi)^3} \frac{~^{(0)}I_n
(p)~^{(1)}I_{n'} (p) e^{-\frac{p^2}{\beta^2_0}}}{\left(\Delta
M^{*} +\omega_\eta+
\frac{\vep^2}{2M_B}+\frac{(\vep-\vek)^2}{2M^*_B}\right)}.\label{14}\ee

\be \Lc^{(2)} =\int \frac{d^3p~p^2}{(2\pi)^3} \frac{~^{(1)}I_n
(p)~^{(0)}I_{n'} (p) e^{-\frac{p^2}{\beta^2_0}}} { \left(\Delta
M^{*} + \frac{\vep^2}{2\tilde M_{BB^*}}\right)}.\label{15}\ee

For $\Lc^{(1)}_s, \Lc^{(2)}_s$  one replaces $\Delta M^*$ with
$\Delta M^*_s$ and $M_B, M_B^*$ with $M_{B_s}, M_{B^*_s}$.  Here
$\beta^{-2}_0=\frac{1}{\Delta_n}+\frac{1}{\Delta_{n'}}$.

The width of the $\Upsilon (n,n') \eta$ decay is obtained  from
$|\Mc|^2$ averaging over vector polarizations as \be \Gamma_\eta
=\frac13 \sum_{i,i'} |\Mc|^2 d\Phi = \frac{2 k^2}{27} \gamma^2
e^{-\frac{k^2}{2\beta^2_2}} d\Phi \left|u_n
\left(\frac{\Lc_s^{(1)}}{\omega^3_s}-\frac{\Lc^{(1)}}{\omega^3}\right)+u_{n'}
\left(\frac{\Lc_s^{(2)}}{\omega^3_s}-\frac{\Lc^{(2)}}{\omega^3}\right)\right|^2\label{16}\ee
where the phase space factor
$d\Phi=\frac{d^3k}{(2\pi)^3}2\pi\delta(M(\Upsilon(n))-
M(\Upsilon(n'))-\omega_\eta-\frac{k^2}{2M(\Upsilon(n'))})$.

Introducing the average $\bar \omega=\frac12 (\omega_s+\omega)$,
one can rewrite (\ref{16}) as $$ \Gamma_\eta=
\left(\frac{M_{br}}{f_\eta}\right)^2\left(\frac{M_{\omega}}{2\bar
\omega}\right)^2\zeta_\eta \frac{k^3}{\bar \omega^4}
e^{-\frac{k^2}{2\beta^2_2}}\left| u_n\left[\left(\frac{\bar
\omega}{\omega_s}\right)^3 \Lc_s^{(1)}- \left(\frac{\bar
\omega}{\omega}\right)^3\Lc^{(1)} \right] +\right.$$ \be\left.
 u_{n'}\left[\left(\frac{\bar
\omega}{\omega_s}\right)^3 \Lc_s^{(2)}- \left(\frac{\bar
\omega}{\omega}\right)^3\Lc^{(2)} \right] \right|^2\label{17}\ee
with $\zeta_\eta=\frac{16}{81\pi N^2_c}\cong 7\cdot 10^{-3}$.

One can see from the general structure of $\Gamma_\eta$, that the
main effect comes from the difference $\left| \left(\frac{\bar
\omega}{\omega_s}\right)^3-\left(\frac{\bar
\omega}{\omega}\right)^3\right|\approx |0.882 -1.139|\approx
0.257$, and  from  the difference of
$|\Lc_s^{(i)}-\Lc^{(i)}|\leq0.05 $.

\section{Results and discussion}

We consider here the single $\eta$ emission in bottomonium
transitions $\Upsilon (n,1) \eta$ with $n=2,3,4,5$. The
corresponding values of $\Delta M^*, \Delta M^*_s, \omega_\eta, k$
are given in the Table 1.
\newpage

{\bf Table 1.}\\ Mass parameters of $\Upsilon (n,n')\eta$
transitions (all in GeV, $k$ in GeV/c) and matrix elements $
\Lc^{(i)}, \Lc^{(i)}_s$ (in GeV).

 \begin{center}
\vspace{3mm}

\begin{tabular}{|l|l|l|l|l|} \hline
&&&&\\

$(n,n')$& 2,1&3,1&4,1&5,1\\
\hline  $\Delta M^*$ & 0.582&0.25&0.026&-0.26\\
\hline  $\Delta M^*_s$ & 0.757&0.425&0.20&-0.08\\
\hline $\omega_\eta$ & 0.562& 0.87&1.075&1.325\\
\hline  $k$& 0.115&0.674&0.923&1.20\\
\hline $\Lc^{(1)}$&
0.263&-0.188&0.29&1.48\ $\cdot 10^{-3}$\\
\hline $\Lc^{(1)}_s$&
0.240&-0.174&0.121&1.41\ $\cdot 10^{-3}$\\
\hline $\Lc^{(2)}$&
0.390&-0.340&0.255&$\begin{array}{l}-0.347\\-i0.0785\end{array}$\\
\hline

$\Lc^{(2)}_s$&
0.341&-0.298&0.226&$\begin{array}{l}+0.0584\\+i0.102\end{array}$\\
\hline
\end{tabular}

\end{center}

The resulting values of $\Gamma_\eta (n, n')$ have been computed
as in (\ref{17}) with $\omega=0.587$ GeV and $\omega_s=0.639$ GeV,
calculated earlier in \cite{14}, see Table 4 of \cite{9}, and $u_n
=\frac{\beta^2_2}{\Delta_n}$ with $\beta_2=0.48$ GeV, and
$\Delta_n$ both fitted to the  realistic wave functions in
\cite{11}, with $\Delta_n = 2.56; 1.54, 1.21, 1.05$ and 1.35 (all
in GeV$^2$) for $ n=1,2,3,4,5$ respectively.

Results of calculations are given in Table 2.

{\bf Table 2.}\\ Values of $\Gamma_\eta(n,n') $  (in keV)
calculated using Eq. (\ref{17}) $ vs$ experimental data
$\Gamma^{\exp}_\eta (n,n')$  (in keV).

 \begin{center}
\vspace{3mm}

\begin{tabular}{|c|c|c|c|c|} \hline
&&&&\\

$(n,n')$& 2,1&3,1&4,1&5,1\\
\hline
$\frac{\Gamma_\eta(n,n')}{\left(\frac{M_{br}}{f_\pi}\right)^2
\left(\frac{M_\omega}{2\bar \omega}\right)^2}$ & 5.0$\cdot
10^{-2}$& 2.9& 1.81&7.04\\\hline

$\Gamma_\eta^{\exp}(n,n')$ &(0.8$\pm0.3)\cdot 10^{-2}$&-&$4.02\pm 0.6$&-\\
&[4]&&[8]&\\
\hline
\end{tabular}

\end{center}

Looking at the Table 2, one can see, that there is an order of
magnitude agreement with experiment. Indeed, the factor
${\left(\frac{M_{br}}{f_\pi}\right)^2 \left(\frac{M_\omega}{2\bar
\omega}\right)^2}$  can be estimated from $\Upsilon (n, n')\pi\pi$
transitions studied in \cite{9}-\cite{13} to be roughly in the
range $[\frac12, 2]$. At the same time $\Gamma_\eta^{\exp}(2,1)$
differs from our calculated value several times, and more accurate
measurements as well as theoretical calculations  are highly
welcome here. Another point, not shown in Table 2, is the old
upper limit \cite{3} on $B_\eta(3,1)$, namely $B(\Upsilon
(3,1)\eta)<2.2\cdot 10^{-3}$\cite{3}, which yields $\Gamma_\eta
(3,1)< (4.5\pm 0.4)\cdot 10^{-2}$ keV and is two orders of
magnitude below our calculated value. Hopefuly new measurements
can  resolve this disagreement. On theoretical side our formulas
(\ref{14})-(\ref{16}) automatically produce the width
$\Gamma_\eta(n,n')$ of the order of $O(1$ keV), for all $(n,1)$
transitions except for (2,1),  where a small phase space factor
$k^3$ gives two orders of magnitude suppression of
$\Gamma_\eta(2,1)$. For the $\Gamma_\eta(5,1)$ one obtains  a 7
keV value, which is however small as compared with the
$\Gamma_{\pi\pi}(5,1)$, the latter being $O(1$ MeV). For
$\Gamma_\eta (4,1)$ and $ {\Gamma_{\pi\pi}(4,1)}$ from \cite{11}
the calculated ratio is $R_{\eta/\pi\pi} \equiv \frac{\Gamma_\eta
(4,1)}{\Gamma_{\pi\pi} (4,1)} \cong 3\left(\frac{M_\omega}{2\bar
\omega}\right)^2 \left(\frac{f_\pi}{M_{br}}\right)^2\approx 3$
which roughly agrees with experimental value
$R^{\exp}_{\eta/\pi\pi}= 2.41\pm 0.40 \pm 0.12$.

To check stability of our results, we have used for  the wave
function of $B_s$  the realistic wave function different from that
of $B$. As a result  one obtains for
$\frac{\Gamma_\eta(n,1)}{\left(\frac{M_{br}}{f_\pi}\right)^2\left(\frac{M_{\omega}}{2\omega}\right)^2}$
the values $(2.74\cdot 10^{-2}$; 1.13; 0.44; 7.3) keV  for
$n=2,3,4,5$ respectively, which should be compared with numbers in
the upper line of Table 2. The same type of sensitivity occurs for
modifications of other wave functions, implying that our results
strictly speaking yield the correct order of magnitude but  not
exact values of $\Gamma_\eta$.

Summarizing, we have calculated  the single $\eta$ production
width $\Gamma_\eta(n,n')$ for  $\Upsilon (n,1) \eta$ transitions
with $ n=2,3,4,5$. We have found that $\Gamma_\eta(n,1)$ are of
the order of and larger than  $\Gamma_{\pi\pi} (n,1)$ for $n=3,4$.
This fact is in agreement with the latest measurements in \cite{8}
of $\Gamma_\eta^{\exp}(4,1)$ and disagrees with earlier
experimental limit on $\Gamma_\eta^{\exp}(3,1)$. Our calculations
do not contain fitting parameters; the only two parameters
$M_{br}, M_\omega$ are fixed by previous comparison with dipion
data.  One should stress that $\eta$ production in bottomonium is
not suppressed in our approach as compared to   $\eta$ production
in charmonium transitions. This  is in contrast with the results
of  method of \cite{6}.

 The financial support of  grants RFFI
06-02-17012,  06-02-17120 and NSh-4961.2008.2 is gratefully
acknowledged.

\vspace{2cm}

{\bf Appendix 1}\\

{\bf Matrix element of single $\eta$ emission}\\

 \setcounter{equation}{0} \def\theequation{A.\arabic{equation}}

According to the general theory in \cite{9,11},  the matrix
elements $\Mc^{(1)}_{\eta}, \Mc_\eta^{(2)}$ for
$\Upsilon(n,n')\eta$ corresponding to diagrams of Fig.1, (a)  and
(b) respectively, can be written as \be \Mc_\eta^{(1)} (n)
=\frac{M_{br} M_\omega}{f_\eta N_c\sqrt{2\omega_\eta}} \int
\frac{d^3p}{(2\pi)^3} \sum_{n_2,n_3} \frac{J_{n
n_2n_3}^{(\vep,\vek)} J^+_{n'n_2n_3}
(\vep)}{E-E_{n_2n_3}(\vep)}.\label{A1}\ee Here $n_2, n_3$ are
channels of intermediate state, with e.g. $n_2 =B, B^*, B_s,
B_s^*, ...,$ we omit indices $n_2, n_3$ and write \be J_n (\vep,
\vek) =\int \bar y^{(\eta)}_{n23} \frac{d^3\veq}{(2\pi)^3} \tilde
\Psi_n (c\vep -\frac{\vek}{2} +\veq) \tilde \psi_{n_2} (\veq)
\tilde \psi_{n_3} (\veq-\vek)\label{A2}\ee where $\tilde \Psi_n,
\tilde \psi_{n_i}$ are momentum space  wave functions of $\Upsilon
(nS)$ and $B (B^*)$ mesons respectively.

The vertex factor $\bar y^{(\eta)}_{123}$ is calculated in the
same way as in \cite{9}, namely from the Dirac trace of the
projection operators for the decay process, in our case this is
$\Upsilon (nS)  \to BB^*\eta$. Identifying the creation operators
as $\bar \psi_b \gamma_i \psi_b, \bar  \psi_b\gamma_5 \psi_n, \bar
\psi_b \gamma_k \psi_n ,  ~ n=u,d,s$ and extracting vertex of
$\eta$ creation from the Lagrangian $ \Delta\Lc =- \int \bar
\psi_n \hat U  M_{br} \psi_n d^4 x$ which gives $i\frac{ M_{br}
\bar \psi_n \gamma_5 \hat \lambda \psi_n}{f_n
\sqrt{2\omega_\eta}}$, with $\hat \lambda=\frac{1}{\sqrt{3}}
\left(\begin{array}{lll} 1&&\\&1&\\&&-2 \end{array}\right)$, one
has for the decay process (cf. Appendix 1 of \cite{9}) \be
G(\Upsilon \to BB^*\eta) =tr [\gamma_i S_b (u,w) \gamma_5 S_{\bar
n} (w, x) \gamma_5S_n (x, w) \gamma_k S_{\bar b}
(w',v)]\label{A3}\ee

As  shown in \cite{9}, appendix 1 and 2, the quark Green's
functions can be split into two factors $S(x,y) =\Lambda^\pm
G(x,y)$, with the projection operators  $\Lambda^\pm_k =\frac{
m_k\pm \omega_k \gamma_4 \mp i p_i ^{(k)} \gamma_i}{2\omega_k},$ $
k=b,n$ and  the scalar part $G(x,y)$, where spins are present only
in spin-depdndent interaction and treated as corrections. Here
$\omega_k$ is the average energy of quark in given meson. Hence
one is brought to the spin factor $Z$. \be Z=tr (\gamma_i
\Lambda^+_b \gamma_5 \Lambda^-_n \gamma_5 \Lambda_n^+ \gamma_k
\Lambda^-_b)\label{A4}\ee
 which is equal to
 \be
 Z=\frac{m^2_b+\Omega^2}{4\Omega^2\omega^2} ((\omega^2
 -\vep^q\vep^{\bar q}) \delta_{ik} - p^q_i p^{\bar q}_k + p^q_k
 p^{\bar q}_i).\label{A5}\ee

 Here  $\Omega, \omega$ are average energies of $b$ and $n$ quark
 in $B$ or $B^*$. One can identify the momenta of $B$ and $B^*$ as
 $\veP_1 =\vep$ and $\veP_2 = -\vep-\vek$, then $\veq$ in
 (\ref{A2}) can be expressed as
 \be \vep_{\bar q} =- \veq +\frac{\omega}{\omega +\Omega} \vep,
 ~~\vep_q = \veq -\frac{\omega}{\omega+\Omega} \vep - \vek
 \frac{\Omega+2\omega}{\Omega+\omega},\label{A6}\ee
 and $Z$ is (we put $m_b \cong\Omega)$
 \be
 Z=\frac{1}{2\omega^2}(-\vek\veq\delta_{ik} +k_iq_k+
 k_kq_i).\label{A7}\ee
 It is important, that we are looking for the $P$-wave of emitted
 $\eta$, and hence for $P$ wave of relative $BB^*$ motion, hence
 the integral (\ref{A2}) should yield the term $\vep\vek$. This
 indeed happens, when one  approximates $\tilde \Psi_n, \tilde
 \psi_n$ as series of oscillator wave functions and (\ref{A2}) has
 the form
 \be
 J_n (\vep, \vek) =\bar y_{n23}^\eta
 e^{-\frac{\vep^2}{\Delta_n}-\frac{\vek^2}{4\beta^2_2}}~^{(0)}I_n
 (\vep).\label{A8}\ee
In the process of $d\veq$ integration in (\ref{A2}) one changes
the integration  variable $q_i \to q'_i - u_np_i +O(k_i)$ with
$u_n =\beta_2^2/\Delta_n$ are  oscillator parameters, found by
$\chi^2$ procedure.

Thus result of $d^3q $ integration yields \be\bar y^\eta_{123} =
\frac{u_n}{2\omega^2} (-\vek\vep \delta_{ik} -k_i p_k+ k_k
p_i).\label{A9}\ee

In an analogous way one obtains for $J_{n'}(\vep)$ in (\ref{A1})
the form \be J_{n'} (\vep) = \bar y_{n'23}^{(\eta)}
e^{-\vep^2/\Delta_{n'}}~^{(1)} I_{n'} (\vep)\label{A10}\ee and
$\bar y^{(\eta)}_{n'23}$ is obtained from the Dirac trace for the
process $B\bar B^*\to \Upsilon (n'S)$, \be Z_2 (BB^*)
=\frac{1}{2\omega} e_{i'kl} (-2q_l +\frac{2\omega}{\omega+\Omega}
p_l)\label{A11}\ee and the result of integration over $d^3q$
yields in (\ref{A10}) \be \bar y_{n'23}^{(\eta)} =- e_{i'kl}
\frac{p_l}{\omega}.\label{A12}\ee

Here $i'$ is the polarization of  $\Upsilon(n'S)$ (represented by
$ \bar\psi_b \gamma_{i'} \psi_b$) and $k$ as in (\ref{A9}) is the
polarization of $B^*$. Averaging over angles of  $\vep$ one
obtains\be \lan \bar y^{(\eta)}_{n23} \bar y^{(\eta)}_{n'23}\ran_p
= \frac{u_n}{3} \frac{\vep^2}{\omega^3} (e_{i'il}
k_l)\label{A13}\ee and finally one writes as in (\ref{12}) \be
\Mc_\eta^{(1)} (n,n') = \frac{M_\omega M_{br} u_n 2 e_{i'il}
k_l}{f_\eta \sqrt{2\omega_\eta} \sqrt{3} \cdot 3  }\left(
\frac{\Lc^{(1)}}{\omega^3} -\frac{\Lc^{(1)}_s}{\omega_s^3}\right)
e^{-\frac{k^2}{4\beta^2_2}}.\label{A14}\ee

For $\Mc^{(2)}_\eta (n, n')$ one can use time inversion and
interchange indices $i,i'$ and change sign of $\vek$, obtaining in
this way Eqs. (\ref{14}) and (\ref{15}) of the main text.  For the
intermediate state of $B^*B^*$ the  summation over polarizations
of $B^*$ yields a net zero result, therefore we are left with only
$(B\bar B^*+ B^* \bar B)$ intermediate state.

\end{document}